\documentclass[a4paper,11pt]{article}

\pdfoutput=1 % if your are submitting a pdflatex (i.e. if you have
\usepackage{jheppub} % for details on the use of the package, please
\usepackage[T1]{fontenc} % if needed

\usepackage{amssymb}
\usepackage{physics}
\usepackage[utf8]{inputenc} %utf8
\usepackage{graphicx}
\usepackage[title,toc,titletoc]{appendix}
\usepackage{braket}
\usepackage[vcentermath]{youngtab}

 \usepackage{fancyhdr}

\usepackage{mathrsfs}
\usepackage[T1]{fontenc}
\usepackage{setspace}
\usepackage{amsfonts}
\usepackage{amssymb}
\usepackage{amsmath}
\usepackage{epsfig}
\usepackage{latexsym}
\usepackage{color}
\usepackage{nicefrac}
 \usepackage{slashed}
 \usepackage{multirow}
 \usepackage{comment}
 \usepackage{soul}
 \usepackage{hyperref}
\usepackage{slashed}
\usepackage{simpler-wick}
\usepackage{array}
\usepackage{tabu}
\usepackage{tcolorbox}
\usepackage{ytableau}
\usepackage{youngtab}
\usepackage{listings}
\usepackage[english]{babel}
\usepackage[utf8]{inputenc}
\usepackage[T1]{fontenc}
\usepackage{bold-extra}
\usepackage{subfig}
\usepackage[subfigure]{tocloft}
\usepackage{comment}
\usepackage{amsfonts, amsmath, amsthm, amssymb}
\usepackage{amscd}
\usepackage{amsfonts}
\usepackage{anonchap}
\usepackage{mathtools}
\usepackage[hang, flushmargin]{footmisc}
\usepackage{footnotebackref}
 \usepackage{float}
\usepackage{physics}
\usepackage{pdfpages}
\usepackage{slashed}
\usepackage[title,toc,titletoc]{appendix}
\usepackage{titlesec}
\usepackage{tocloft}
\usepackage{bbm}
\usepackage{bbold}
\usepackage{tikz-cd}
\definecolor{rossoCP3}{cmyk}{0,.88,.77,.40}
\usetikzlibrary{decorations.pathmorphing}

\newcommand{\ea}[1]{
\begin{align}
#1
\end{align}
}

\newcommand{\ee}{\end{equation}}
\newcommand{\be}{\begin{equation}}
\newcommand{\bea}{\begin{align}}
\newcommand{\eea}{\end{alig}}

\newcommand   \cO {\mathcal{O}}

\newcommand{\identity}{\mathbbm{1}}

  \definecolor{darkrosso}{RGB}{100,13,20}
\definecolor{lightor}{RGB}{248,150,30}

\def\lsim{\mathrel{\rlap{\lower4pt\hbox{\hskip1pt$\sim$}}
    \raise1pt\hbox{$<$}}}                % less than or approx. symbol
\def\gsim{\mathrel{\rlap{\lower4pt\hbox{\hskip1pt$\sim$}}
    \raise1pt\hbox{$>$}}}                % greater than or approx. symbol
    
    \def\be{\begin{equation}}
\def\ee{\end{equation}}
\def\ba{\begin{eqnarray}}
\def\ea{\end{eqnarray}}

\def\IR{\relax{\rm I\kern-.18em R}}

\def\IR{\relax{\rm I\kern-.18em R}}
\def\IL{\relax{\rm I\kern-.18em L}}

\def\inv{^{\raise.15ex\hbox{${\scriptscriptstyle -}$}\kern-.05em 1}}

\def\cO{{\cal O}}

\def\Tr{{\rm Tr}}

\title{ \centering Scaling results \\for \\ Charged sectors of near conformal QCD}

\author[a]{Jahmall Bersini,}
\author[b, c, e]{Alessandra D'Alise,}
\author[d, e]{Clelia Gambardella, }
\author[b, c, d, e]{Francesco Sannino,}

\affiliation[a]{ Kavli IPMU (WPI), UTIAS, The University of Tokyo, Kashiwa, Chiba 277-8583, Japan}
\affiliation[b]{Dipartimento di Fisica ``E. Pancini", Università di Napoli Federico II, via Cintia, 80126 Napoli, Italy}
\affiliation[c]{INFN sezione di Napoli, Complesso Universitario di Monte S. Angelo Edificio 6, via Cintia, 80126 Napoli, Italy}
\affiliation[d]{Scuola Superiore Meridionale, Largo S. Marcellino, 10, 80138 Napoli NA, Italy}
\affiliation[e]{Quantum  Theory Center ($\hbar$QTC) at IMADA and D-IAS, Southern Denmark Univ., Campusvej 55, 5230 Odense M, Denmark}

\emailAdd{jahmall.bersini@ipmu.jp}
\emailAdd{alessandra.dalise@unina.it}
\emailAdd{c.gambardella@ssmeridionale.it}
\emailAdd{sannino@qtc.sdu.dk}

\abstract{ We provide the leading near conformal corrections on a cylinder to the scaling dimension $\Delta_Q^\ast$ of fixed isospin charge $Q$ operators  defined at the lower boundary of the Quantum Chromodynamics conformal window:  \begin{equation}
     \Delta_Q =  \Delta_Q^\ast  +\left(\frac{m_{\sigma}}{4 \pi \nu}\right)^2 \, Q^{\frac{\Delta}{3}} B_1 +  \left(\frac{m_\pi(\theta)}{4\pi \nu} \right)^4\  Q^{\frac{2}{3}(1-\gamma)} B_2  + \mathcal{O}\left ( m_\sigma^4 , m_\pi^8, m_\sigma^2 m_\pi^4\right)  \ . \nonumber
     \end{equation}
The results are expressed in powers of the dilaton and pion masses in units of the chiral symmetry breaking scale $4\pi \nu$ with the theta-angle dependence encoded directly in the pion mass.  The characteristic $Q$-scaling is dictated by the quark mass operator anomalous dimension $\gamma$ and the one characterising the dilaton potential $\Delta$. The coefficients $B_i$ with $i=1,2$ depend on the geometry of the cylinder and properties of the nearby conformal field theory.  
}

\date{}
\begin{document}
% \maketitle\let\thefootnote\relax\footnotetext
\maketitle

\section{Introduction}
\label{intro}
Unveiling near-conformal properties of Quantum Chromodynamics (QCD) has attracted much interest over the past several decades. This exploration was spurred by the seminal work of Banks and Zaks \cite{Banks:1981nn} that discovered the existence of a perturbative infrared fixed (IR) point for massless QCD for a number of flavours $N_f$ just below the loss of asymptotic freedom. As one decreases $N_f$ relative to the fixed number of colours $N_c$ one expects below a critical number of flavours $N_f^c$ the theory to undergo a quantum phase transition. That this transition is bound to occur is clear from the fact that for the observed  number of light flavours the theory breaks chiral symmetry dynamically generating a non-perturbative scale even in the absence of explicit light quark masses. Estimates of the critical number of flavours $N_f^c$ are all non-perturbative in nature and indicate that this value is, for three colours QCD,  around $N_f^c \sim 10 - 12$. The window in the flavour-colour space where the theory displays IR conformality is termed conformal window and its phase diagram for arbitrary matter representations is presented in \cite{Sannino:2004qp,Dietrich:2006cm}. Ongoing lattice studies are continuously searching to elucidate the confines and properties of the conformal window as summarised in \cite{Cacciapaglia:2020kgq}. 

Several key questions are related to the dynamics near the lower boundary of the conformal window. These range from a precise determination of its lower edge to the characterisation of the quantum phase transition. One exciting possibility is for the transition to lose conformality a la Berezinskii-Kosterlitz-Thouless (BKT) \cite{Berezinsky:1970fr,Kosterlitz:1973xp,Kosterlitz:1974nba}. The latter occurs in two dimensions and was envisioned for four dimensions in \cite{Miransky:1996pd, Miransky:1984ef,Appelquist:1996dq,Gies:2005as}. An alternative is for the quantum transition to be a jumping one \cite{Sannino:2012wy}. The first possibility leads to an infrared non-conformal physics displaying premonitory signs of near-conformality that can be observed in power-law scaling of certain phenomenologically relevant operators \cite{Holdom:1988gs, Holdom:1988gr,Cohen:1988sq}. This dynamics is also known as walking dynamics since the underlying gauge coupling has a region in the renormalization time scale where the coupling is almost constant, and therefore it walks rather than displaying a running behaviour. Such a walking behaviour has recently been shown to  mathematically describe the endemic state of pandemics \cite{Complex:2021} via the epidemic renormalization group approach \cite{DellaMorte:2020wlc}. This methodology was also used to successfully predict the second wave COVID-19 pandemic in Europe \cite{Secondwave:2021} providing policymakers crucial epidemiological information.  In terms of the spectrum of the theory, in the confining and chiral symmetry broken phase occurring below but near the lower end of the conformal, besides the ordinary Goldstons, it has long been argued \cite{Leung:1985sn,Bardeen:1985sm,Yamawaki:1985zg,Sannino:1999qe} that another precursor of a smooth quantum phase transition is the occurrence of a dilaton in the theory. Its description at the effective Lagrangian level goes back to the work of Coleman in \cite{coleman1988aspects} and for walking dynamics considered in \cite{Sannino:1999qe,Hong:2004td,Dietrich:2006cm, Dietrich:2005jn,Appelquist:2010gy}. Recent investigations via effective approaches in different dynamical regimes have appeared in \cite{Chacko:2012sy, Matsuzaki:2013eva, Hansen:2016fri, Golterman:2016lsd, Cata:2019edh, Golterman:2020tdq, Appelquist:2020bqj, Appelquist:2022mjb, Cata:2018wzl, appelquist2020dilaton, Orlando:2020yii, Bersini:2022bnx, Bersini:2023uat}. An explicit calculable example has been discussed in  \cite{Antipin:2011aa} where it was possible to demonstrate the emergence of a dilaton in a near-conformal field theory (CFT) alongside a precise study of all the relevant scales emerging once conformality is lost. A complementary analysis of the mass-induced confinement for gauge-fermion theories near the lower edge of the conformal window was performed in \cite{Marcarelli:2022sbb}. One can also use first principle lattice simulations to disentangle the dilaton properties, however, this is a difficult task since its quantum numbers are the ones of the vacuum and therefore it is subject to large numerical noise. Nevertheless there are attempts to fit effective approaches to lattice data as summarized in  \cite{Ingoldby:2023mtf}).

Complementary ways to isolate the dilaton properties and more generally to learn about the near-conformal dynamics of QCD are therefore vital to corner the properties of the flavour-driven quantum phase transition.     As we shall see fixed charged sectors offer novel opportunities to disentangle near conformal dynamics providing precious information on the sectors responsible for breaking conformality.  We start by recalling that central quantities in any CFT are scaling dimensions of local operators.  By the state-operator correspondence \cite{Cardy:1984rp} these are the energies of the corresponding states on a non-trivial gravitational background. For example, the scaling dimension of the lowest scaling operator of charge $Q$ denoted with $\Delta_Q^\ast$ is mapped into ground state energy $E_Q$  on the cylinder   via the relation 
\begin{equation}
    \Delta_Q^\ast = r \, E_Q \ ,
\end{equation}
with $r$ the radius of the cylinder. In the large charge limit we can compute scaling dimensions of fixed charge operators by means of semiclassical computations \cite{Hellerman:2015nra, Banerjee:2017fcx, Orlando:2019skh, Gaume:2020bmp, Hellerman:2023myh, Antipin:2020abu, Antipin:2022naw, Antipin:2022hfe, Antipin:2023tar, Monin:2016jmo, Badel:2019oxl, Cuomo:2020rgt, Arias-Tamargo:2019xld}. Following \cite{Orlando:2019skh, Orlando:2020yii, Bersini:2022bnx} it is natural to extend this relation to near-conformal field theories by introducing the quantity: 
\begin{equation}
    \Delta_Q \equiv r \, E_Q  =  \Delta_Q^\ast  + {\rm near~CFT~terms}\ . 
\end{equation}
The near CFT terms depend on the way the CFT is deformed. In the case at hand we have two sources of conformal breaking: One stemming from an explicit quark mass term and the other from the occurrence of an operator of dimension $\Delta$ inducing a dilaton-potential.  Therefore, in this work, we extend the chiral Lagrangian to include a dilaton sector and use the large charge expansion framework to determine the general expression for $\Delta_Q$ arriving at one of our central results: 
 \begin{equation}
 \label{Alessandra}
     \Delta_Q =  \Delta_Q^\ast  +\left(\frac{m_{\sigma}}{4 \pi \nu}\right)^2 \, Q^{\frac{\Delta}{3}} B_1 +  \left(\frac{m_\pi(\theta)}{4\pi \nu} \right)^4\  Q^{\frac{2}{3}(1-\gamma)} B_2  + \mathcal{O}\left ( m_\sigma^4 , m_\pi^8, m_\sigma^2 m_\pi^4\right)  \ . 
     \end{equation}
The latter is expressed in powers of the dilaton and pion masses given in units of the chiral symmetry breaking scale $4\pi \nu$. The theta-angle dependence is explicitly encoded in the pion mass.  The novel scalings in the charge corrections are expressed in terms of the quark mass operator anomalous dimension $\gamma$ and the one characterising the dilaton potential $\Delta$. The coefficients $B_i$ with $i=1,2$ depend on the geometry of the cylinder and the properties of the nearby CFT.  The above result for $\Delta_Q$ in \eqref{Alessandra} is obtained as the large charge limit of the general expression given in \eqref{deltagenerale} at leading order in the semiclassical expansion. The next-to-leading order, to be computed in the future, requires the knowledge of the spectrum of fluctuations that we have also determined here.  Additionally, we have also provided the phase diagram of QCD at non-zero isospin chemical potential in the presence of the CP-violating topological term.

The paper is organised as follows. In Section~\ref{dilaton} we introduce the QCD chiral Lagrangian for generic $N_f$ including the $\theta$-angle and isospin chemical potential $\mu$. The $\mu-\theta$ phase diagram is presented in Sec.~\ref{phase}. The dilaton potential and setup for the large charge expansion are discussed in Sec.~\ref{DACL}. Section \ref{sectiongse} is devoted to the determination of $\Delta_Q$ while in Sec.~\ref{sectionfluc}  we first discuss the patterns of symmetry breaking and then compute the spectrum of fluctuations. Section~\ref{Casimiren} is dedicated to the universal contributions in the conformal limit. We offer our conclusions in Sec.~\ref{Conclusions}.

\section{Chiral Lagrangian at finite isospin and $\theta$-angle: Notation and conventions}
 \label{dilaton}

The low-energy dynamics of QCD at finite isospin density is described by the chiral Lagrangian below
\begin{align}
\label{lagtheta}
    \begin{split}
    \mathcal{L} &= \nu^2 Tr\{ \partial_\mu\Sigma\partial^\mu\Sigma^\dagger\}+m^2_\pi \nu^2 Tr\{M\Sigma+M^\dagger\Sigma^\dagger\} + 2i\mu \nu^2 Tr\{I\partial_0 \Sigma \Sigma^\dagger - I\Sigma^\dagger \partial_0 \Sigma \}  \\& + 2\mu^2 \nu^2 Tr\{II - \Sigma^\dagger I\Sigma I\}  \ , 
    \end{split}
\end{align}
where $4 \pi \nu$ is the energy scale of the chiral symmetry breaking, $\mu$ is the isospin chemical potential, and
\begin{equation}
    \Sigma = e^{i \varphi/\nu} \,, \qquad \varphi = \Pi^a T^a + \frac{S}{\sqrt{N_f}} \ ,
\end{equation}
with $T^a$ the $SU(N_f)$ generators normalized as $\Tr [T^a T^b ] = \delta^{ab}$.
The mass matrix and the isospin generator read
\begin{equation}
\label{Massa}
M = \identity_{N_f} \,, \qquad I = \frac{1}{2} \left(\begin{array}{cc}
\identity_{N_f/2} & 0  \\ 0 & -\identity_{N_f/2}
\end{array}\right)  \ ,
%\equiv \frac{1}{2}  \tilde{\tau}_3
\end{equation}
where we assumed degenerate Goldstone bosons of mass $m_{\pi}$. The matrix $I$ generalizes the concept of isospin in the multi-flavor theory with the two flavour case being the conventional QCD isospin. 

Finally, as fully detailed in \cite{Witten:1980sp, DiVecchia:1980yfw}, the CP-violating topological sector is included through the $\theta$-angle term
\be
\Delta\mathcal{L}_{\theta} =  -a \nu^2\left(\theta-\frac{i}{2}Tr\{\log \Sigma - \log \Sigma^\dagger \}\right)^2 \ ,
\ee
where $a$ is the energy scale associated with the axial anomaly.

\section{Phase diagram in the $\mu$-$\theta$ plane}
\label{phase}

In the absence of the $\theta$ angle, the ground state of our theory generalizes the $N_f=2$ case examined in detail in \cite{Son:2000xc} and takes the following form 
\begin{align}
\label{eq:ansatsvacuum}
 \Sigma_c &=  \identity_{N_f} \cos \varphi + i \Sigma_I \sin\varphi\ ,
\end{align}
with
\begin{align}
\Sigma_I&= \left(\begin{array}{cc} 0 & \identity_{N_f/2}  \\ \identity_{N_f/2} &  0 \\
\end{array} \right)\ \cos\eta\  
+i \left(\begin{array}{cc} 0 & -\identity_{N_f/2}  \\ \identity_{N_f/2} &  0  \end{array} \right)\ \sin\eta \, . 
\end{align}
The two terms in $\Sigma_c$ minimize, respectively, the mass and isospin terms in the potential. In fact, it can be easily shown that the latter is minimized when $\Sigma_I$ satisfies $I \Sigma_I = - \Sigma_I I$. To take into account the effect of the $\theta$-angle on the vacuum state we introduce the Witten variables $\alpha_i$ \cite{Witten:1980sp} and arrive at our ansatz for the ground state 
\be
\label{eq:ansatsvacuumcomplete}
\Sigma_0 = U(\alpha_i) \Sigma_c\ , \,\, \text{with}\  \, \, U(\alpha_i) = \text{diag}\{e^{-i \alpha_1}\,, \dots \,,e^{-i \alpha_{N_f}} \}\ .
\ee
It is useful to define the following quantities
\begin{equation}
    \Bar{\theta}=\theta-\sum_{i}^{N_f}\alpha_i\ ,  \,\,\, X=\sum_{i=1}^{N_f} \cos\alpha_i\ ,
\end{equation}
in terms of which the Lagrangian of the theory evaluated on the ground state ansatz reads
\begin{equation}
    \mathcal{L}[\Sigma_0]=2 m_\pi^2 \nu^2 X \cos \varphi +N_f\mu ^2 \nu^2 \sin ^2 \varphi -a \nu^2 \Bar{\theta}^2 \,.
\end{equation}
The angle $\varphi$ and the Witten variables $\alpha_i$ are determined by the equations of motion (EOMs) as
\begin{align}\label{primaeq}
    \sin \varphi  \left(N_f\cos \varphi -\frac{m_\pi^2 X}{\mu ^2}\right)&=0 \ , \\ \label{thetavac}
     m_\pi^2 \sin \alpha_i \cos \varphi &=  a \Bar{\theta}\ , \,\,\, i=1,\dots, N_f \ ,  
\end{align}
while the angle $\eta$ does not appear in the equation of motion meaning that there is a residual unbroken $U(1)$ isospin vector-symmetry stemming from the original diagonal $SU(2)$ isospin. However, once we choose a specific value of $\eta$ this amounts to the further spontaneous breaking of this leftover $U(1)$.  The first EOM has two solutions, namely $\varphi= 0$ and $\cos\varphi =\frac{m_\pi^2 X}{N_f\mu^2}$. When the latter is realized the theory is in a superfluid phase characterized by pion condensation. The energy of the system in the two phases reads
\begin{align} 
    E(\theta)&=-2 m_\pi^2 \nu^2 X+a\nu^2 \Bar{\theta}^2\ \,\,\,\,\, &\text{normal phase} \nonumber \ & \ \ (\varphi=0)\\\label{energiathetasup}
    E(\theta)&=-\frac{m_\pi^4\nu^2}{N_f\mu^2} X^2-N_f\nu^2\mu^2+a\nu^2 \Bar{\theta}^2\ \,\,\,\,\, &\text{superfluid phase}&\ \left(\cos\varphi =\frac{m_\pi^2 X}{N_f\mu^2}\right)\ .
\end{align}
We first observe that when $a\gg m^2_\pi$ all the $\theta$-dependence can be read off the effective pion mass $m^2_\pi(\theta)=m^2_\pi X/N_f$. The EOMs and the expression for the vacuum energy of the system are very similar to the ones found in \cite{Bersini:2022jhs} for two-color QCD with finite baryon density. 
At $\theta=0$ the normal to superfluid phase transition occurs at a critical value of the chemical potential $\mu_c = m_\pi$. Since the $\theta$-vacuum may differ in the two phases, to study the superfluid transition at non-vanishing values of $\theta$, we first need to determine the $\theta$ dependence.
We then solve eq.\eqref{thetavac} by expanding in powers of $m^2_\pi/a$ that we take to be small. Specifically, at the leading order in $m^2_\pi/a$, we have
\begin{align} \label{Solgen2}
    \alpha_{i}=\begin{cases}
  \pi- \alpha(\theta) \,,\qquad  & i=1,\dots,n \\
 \alpha(\theta) \,,\qquad & i=n+1,\dots,N_f \,,
    \end{cases} 
\end{align}
where
\begin{equation} \label{Solgen}
   \alpha(\theta)=\frac{\theta+ (2k-n)\pi}{(N_f-2n)},\quad k=0,\dots, N_f-2n-1 \,, \quad n=
    0,..., \left[\frac{N_f-1}{2}\right] \,.
\end{equation}
The interval of values for $k$ is constrained because at fixed $n$ the solutions are periodic in $k$ of period $N_f-2n$. In the normal phase the energy is minimized when $X$ is maximized. Hence, it can be shown that the $\theta$-vacuum corresponds to $n=0$ and the following values of $\alpha(\theta)$
\be \label{questaequazione}
\alpha(\theta)=\begin{cases}
    \frac{\theta}{N_f}  \,, \qquad \theta \in [0, \pi] \\
     \frac{\theta -2\pi}{N_f}\,, \ \quad \theta \in [\pi, 2\pi]  \,.
\end{cases} 
\ee

\begin{figure}[h!]
    \centering
    \subfloat[\centering $\theta$-dependence of the energy in the normal phase for $N_f=2$.]{{\includegraphics[scale=0.5]{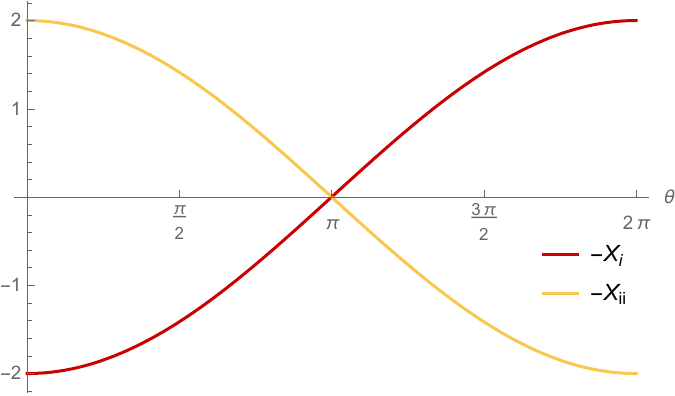} }}%
    \qquad
    \subfloat[\centering $CP$ order parameter for $N_f=2$.]{{\includegraphics[scale=0.5]{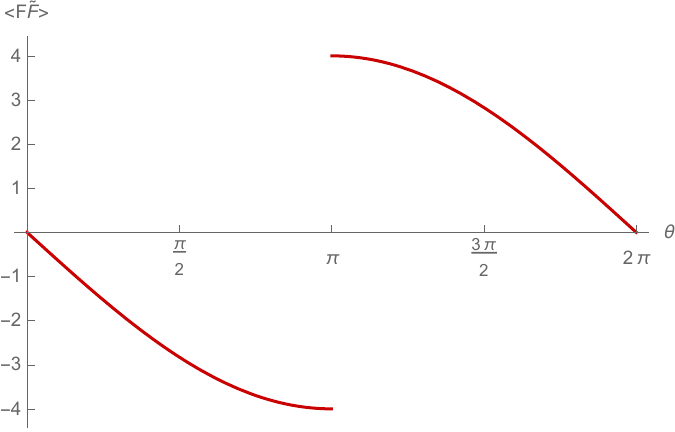} }}%
    \caption{$\theta$-dependence of the normalized ground state energy and $CP$ order parameter $\langle F \widetilde{F}  \rangle$ as a function of $\theta$ for $N_f=2$ in the normal phase. Here $X_i = 2 \cos\left(\theta/2 \right)$ and  $X_{ii} = 2 \cos\left(\theta/2 - \pi \right)$ refers to the two solutions for $\alpha(\theta)$ in eq.\eqref{questaequazione}.  }
    \label{fig:minima}%
\end{figure}

The physics at $\theta=\pi$ deserves further discussion  In fact, the Lagrangian possesses $CP$ symmetry when $\theta=\pi$ but the latter is spontaneously broken by the vacuum leading to the known Dashen's phenomenon \cite{Dashen:1970et}. In fact, at $\theta=\pi$ the two solutions for $\alpha(\theta)$ in eq.\eqref{questaequazione} have the same energy leading to degenerate vacua connected by a CP transformation. In Fig.\ref{fig:minima} we visualize the ground state energy as a function of $\theta$ for the template case $N_f=2$ as well as a plot of the  $CP$ order parameter $\langle \widetilde{F} F \rangle$ associated with the pseudo-scalar glueball condensate.

\begin{figure}[h!]
    \centering
    \subfloat[\centering $\theta$-dependence of the energy in the superfluid phase for $N_f=2$.]{{\includegraphics[scale=0.5]{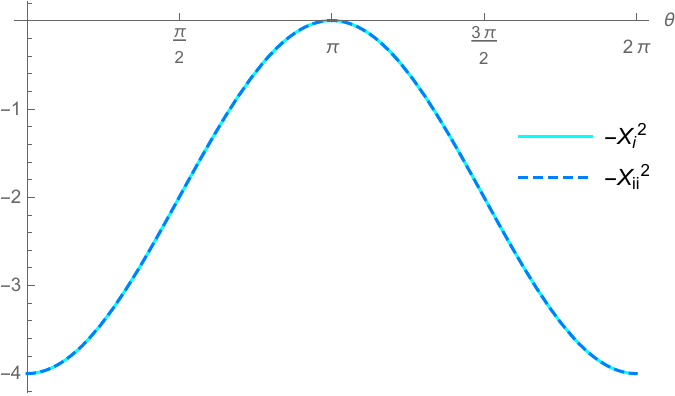} }}%
    \qquad
    \subfloat[\centering $CP$ order parameter for $N_f=2$.]{{\includegraphics[scale=0.5]{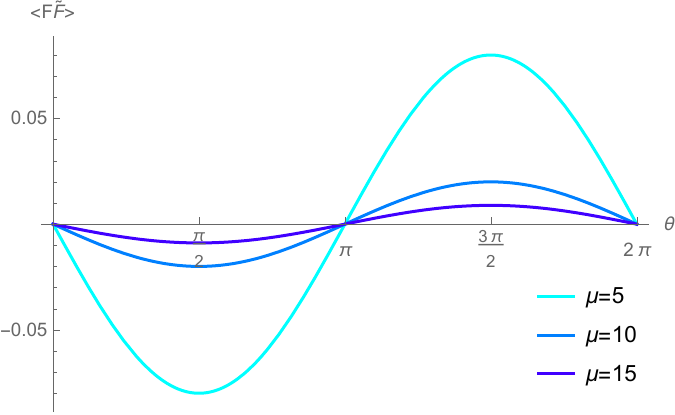} }}%
    \caption{$\theta$-dependence of the normalized ground state energy and $CP$ order parameter $\langle F \widetilde{F}  \rangle$ as a function of $\theta$ for $N_f=2$ in the superfluid phase. Here $X_i^2 = 4 \cos^2\left(\theta/2 \right)$ and  $X_{ii}^2 = 4 \cos^2\left(\theta/2 - \pi \right)$ refers to the two solutions for $\alpha(\theta)$ in eq.\eqref{questaequazione} while $\mu$ is measured in units of $m_\pi$.}%
    \label{fig:minimasup}%
\end{figure}

We now move to the $\theta$-dependence in the superfluid phase. The EOM becomes
\begin{align} \label{superEOM}
\frac{m^4_\pi}{N_f \mu^2}X   \sin \alpha_i &= a \bar{\theta} \ , \,\,\, i=1,\dots,  N_f \,,
\end{align}
and admits the same solution \eqref{Solgen} at the leading order in the natural expansion parameter $\frac{m^4_\pi}{\mu^2}$. The crucial difference with respect to the normal phase is that the energy now depends quadratically instead of linearly on $X$. The situation is analogous to the case of two-color QCD at finite baryon charge investigated in detail in \cite{Bersini:2022jhs} for even $N_f$. As has been found there, the ground state solution is the same as the normal phase being given by eq.\eqref{questaequazione}. Accordingly, for $N_f > 2$, at the crossing point $\theta=\pi$ spontaneous breaking of $CP$ occurs. On the other hand, for $N_f=2$ the two solutions $\alpha(\theta) = \theta/N_f$ and $\alpha(\theta) = \frac{\theta -2\pi}{N_f}$ have degenerate energy for all values of $\theta$. As a consequence, the physics is analytic at $\theta=\pi$, and Dashen's phenomenon does not occur \cite{Bersini:2022jhs, Metlitski:2005db}. The $\theta$ vacuum in the superfluid phase in the two flavour case is illustrated in Fig.\ref{fig:minimasup}. 

Note that the pairs of completely degenerate solutions remain such to all orders in $\frac{m_\pi^2}{a}$. In fact, given the EOM \eqref{superEOM} for a certain $\alpha(\theta)$
\begin{equation}
\frac{m_\pi^4}{2 a \mu ^2}\sin (2 \alpha(\theta) ) = \theta - N_f \alpha(\theta)  \,,
\end{equation}
we have the same EOM for $\alpha(\theta) + \pi$, upon shifting the $\theta$-angle as $\theta \to \theta + N_f \pi$. Being $N_f$ even, this corresponds to a shift by an integer multiple of $2 \pi$ which leaves the physics unaltered. 
Finally, we investigate the critical value of the isospin chemical potential where the superfluid phase transition occurs. For $m_\pi^2 \ll a$, the latter occurs at a critical value of the chemical potential given by
\be
\mu_c = m_\pi(\theta) = m_\pi \left[\sqrt{\abs{\cos \frac{\theta}{N_f}}}+ \mathcal{O}\left(\frac{m_\pi^2}{a} \right)  \right] \,,
\ee
which is of particular interest when $N_f=2$ at $\theta=\pi$. In fact, $\mu_c$ is almost vanishing since the effective pion mass $m^2_\pi (\theta)\sim m^2_\pi\abs{\cos(\theta/2)}$ identically vanishes. Concretely, at $\theta \sim \pi$ we have
\be
\mu_c \sim m_\pi \sqrt{\frac{m_\pi^2}{a} + \frac{\abs{\phi}}{2}} \,, \qquad \phi \equiv \theta - \pi \,.
\ee
On the other hand, a vanishing pion mass at the effective Lagrangian level signals incongruities since it would imply no explicit breaking of chiral symmetry. However, there is no chiral symmetry restoration in the fundamental QCD Lagrangian. This apparent paradox is solved by pointing out that the global flavour symmetry is still broken when including higher-order mass terms in the effective Lagrangian \cite{Smilga:1998dh, Tytgat:1999yx}.

\section{Dilaton augmented chiral Lagrangian and the large charge expansion} \label{DACL}

In this section, following \cite{Orlando:2019skh} we consider the dynamics near the lower edge of the conformal window to determine the ground-state energy on non-trivial metrical backgrounds that can be associated with scaling dimensions of QCD operators carrying (generalized) isospin charge.  The first step is to upgrade the chiral Lagrangian to a conformally invariant theory via the introduction of a scalar degree of freedom $\sigma$, the \emph{dilaton}, which under scale dilations $x\mapsto e^{\lambda} x$ transforms as
\begin{equation}
    \sigma \mapsto \sigma-\frac{\lambda}{f}\ .
\end{equation}
Scale invariance can then be enforced at the effective action level by coupling $\sigma$ to each operator $\mathcal{O}_k$ of dimension $k$ appearing in the Lagrangian as \cite{coleman1988aspects, goldberger2007light}
\begin{equation}
    \mathcal{O}_k \mapsto e^{(k-4)\sigma f}\ \mathcal{O}_k\ .
\end{equation}
The resulting theory features non-linearly realized dilation invariance with $f$ and $\sigma$ being the length scale and the Goldstone boson associated with the spontaneous breaking of conformal symmetry, respectively. Explicit breaking of the latter can be taken into account introducing a potential term for $\sigma$. Consider perturbing a CFT with an operator $\mathcal{O}$ with conformal dimension $\Delta$, i.e. 
\begin{equation}
\mathcal{L}_\text{CFT} \rightarrow \mathcal{L}_\text{CFT}  + \lambda_{\mathcal{O}} \mathcal{O} \,,
\end{equation}
with $\lambda_\mathcal{O}$ the corresponding coupling. For $\lambda_\mathcal{O} \ll 1$ the perturbation generates the following dilaton potential \cite{rattazzi2001comments,goldberger2007light,Chacko:2012sy} 
\begin{equation}\label{sigmapotDelta}
  V(\sigma) =  \frac{m_{\sigma}^2 e^{-4 f \sigma } }{4 (4-\Delta ) f^2} - \frac{m_{\sigma}^2 e^{-\Delta f \sigma }}{\Delta (4-\Delta ) f^2} + \mathcal{O}( \lambda_{\mathcal{O}}^2) \ .
\end{equation}
This potential has been considered in various recent studies of QCD-like theories in the near-conformal regime \cite{appelquist2020dilaton, Appelquist:2022mjb, Bersini:2022bnx} as well as in the description of dense skyrmion matter \cite{Li:2016uzn, Ma:2021nuf, Bersini:2023uat}. For $\Delta = 2$ it reduces to the well-known Higgs-like quartic potential. Note that the potential \eqref{sigmapotDelta} cannot be trusted for $\Delta \ll 1$ and its singularity at $\Delta = 0$ is unphysical \cite{Bersini:2023uat}. When the perturbation is nearly marginal, i.e. $\Delta \to 4$, for arbitrary values of $\lambda_\mathcal{O}$ the potential reads \cite{goldberger2007light} 
\begin{equation} \label{logpot}
  V(\sigma) = -\frac{m_{\sigma}^2 e^{-4 f \sigma }}{16 f^2} (1+4 f \sigma ) +\mathcal{O}\left((\Delta-4)^2\right)  \ .
\end{equation}
The above agrees with the potential derived according to the counting scheme proposed in \cite{Golterman:2016lsd, Golterman:2020tdq}. Since eq.\eqref{logpot} is the $\Delta \to 4$ limit of eq.\eqref{sigmapotDelta}, in what follows we will assume the generic form of the dilaton potential eq.\eqref{sigmapotDelta}. 

Finally, the mass term operator has dimension $y=3-\gamma$, with $\gamma$ being the anomalous dimension of the chiral condensate. Unitarity dictates $0<\gamma<2$ whereas the underlying four fermion operator becomes near-marginal for $\gamma \simeq 1$. For these reasons, this work focuses on the range $0<\gamma<1$. 

Our second step is to employ the large charge expansion framework \cite{Hellerman:2015nra, Badel:2019oxl, Gaume:2020bmp} to determine the scaling dimension $\Delta_Q$ of the lowest-lying operator with charge $Q$. To this end, we exploit the approximate Weyl invariance of the near-conformal theory to map the latter onto the cylinder $\mathbb{R}\times S^{3}$. We denote the volume, the radius, and the Ricci scalar of $S^{3}$ as $V$, $r$, and $R=\frac{6}{r^2}$ respectively. The advantage is that we can now consider an approximate state-operator correspondence which links $\Delta_Q$ to the ground state energy $ E_Q$ at fixed $Q$ of the theory on the cylinder as
\be
\Delta_Q =r E_Q\,, \qquad E_Q=\mu Q - \mathcal{L} \ ,
\ee
The dilaton-pion effective Lagrangian on $\mathbb{R}\times S^3$ reads
{\small
\begin{equation}
\label{lagdressed}
    \begin{split}
    \mathcal{L}_{\sigma}&= \nu^2 Tr\{ \partial_\mu\Sigma\partial^\mu\Sigma^\dagger\}\ e^{-2\sigma f}+ m^2_\pi \nu^2 Tr\{M\Sigma+M^\dagger\Sigma^\dagger\}\  e^{-y\sigma f} +2\mu^2 \nu^2 Tr\{II-\Sigma^\dagger I\Sigma I\} e^{-2\sigma f}\\& + 2i\mu \nu^2 Tr\{I\partial_0 \Sigma \Sigma^\dagger - I\Sigma^\dagger \partial_0 \Sigma \}e^{-2\sigma f}   -a \nu^2\left(\theta-\frac{i}{2}Tr\{\log \Sigma - \log \Sigma^\dagger \}\right)^2\ e^{-4\sigma f} -\Lambda^4_0\ e^{-4\sigma f}  \\& + \frac{1}{2}\left(\partial_\mu\sigma\partial^\mu\sigma-\frac{R}{6f^2}\right)\ e^{-2\sigma f}-V(\sigma) \,,
\end{split}
\end{equation}}
where for later convenience we included the bare cosmological constant $\Lambda_0$. In the conformal limit $m_\pi = m_\sigma = 0$, the ground state energy $E_Q$ can then be computed via a semiclassical expansion in the double scaling limit 
\be \label{dsl}
\Lambda_0 f \to 0 \ , \quad Q \to \infty \ , \quad Q ( \Lambda_0 f)^4 =\text{fixed.}
\ee
This can be seen by considering the expectation value of the evolution operator $U=e^{-H T}$ in an arbitrary state $\ket{Q}$ with charge $Q$
\be
\braket{U}_Q \equiv \braket{Q| e^{-H T} |Q} \underset{T \to \infty}{\to} \mathcal{N} e^{-E_Q T} \,,
\ee
with $H$ the Hamiltonian, $T$ the time interval, and $\mathcal{N}$ an unimportant normalization factor.
Then one can rescale the fields as $\Sigma \to \nu f \Sigma$ and $e^{-f \sigma} \to \sqrt{Q} e^{-f \sigma}$ to exhibit $Q$ as a new counting parameter in the path integral expression for $\braket{U}_Q $. 
% \be
% \braket{U}_Q  = \frac{1}{\mathcal{Z}} \int \mathcal{D }\Sigma\mathcal{D \sigma} \exp \left[  \int^{T/2}_{-T/2}  \int_{S^3} \left(\mu \frac{Q}{V} -\mathcal{L}_{\theta, \sigma} \right) \right]
% \ee
% with $\mathcal{Z}$ the partition function and $\mathcal{L}_{\theta, \sigma}$ the zero matter density Lagrangian. 
Accordingly, the scaling dimension of the lowest-lying operator assumes the following form
\be
E_{Q} R=\Delta_{Q} = \sum_{j=-1} \frac{1}{Q^j}\Delta_j \left(Q (\Lambda_0 f)^4\right) \ .
\label{ERexpansion}
\ee

The leading order $\Delta_{-1}$ is given by the classical ground state energy on $\mathbb{R}\times S^3$ while the next-to-leading order $\Delta_0$ is determined by the fluctuations around the classical trajectory. We refer the reader interested in the details of the approach to \cite{Badel:2019oxl, Antipin:2020abu, Antipin:2022naw}. In the next section, we will determine the classical ground state energy in the large $Q \gg  (\Lambda_0 f)^4$ limit. As we shall see, deviations from conformality will be encoded in a set of contributions that depend on the spacetime geometry due to the lack of Weyl invariance.

% where quadratic terms independent of the fields have been included as well as the bare cosmological constant $\Lamda_0$.
%matias

\section{Large charge expansion: leading order}
\label{sectiongse}
As anticipated in the previous section, the state-operator correspondence
%, realised through the Weyl map, 
enables us to deduce the scaling dimension for the lowest-lying operator with (generalised) isospin charge $Q$. 
This is achieved by determining the energy associated with the vacuum structure inducing the superfluid phase transition.
We therefore evaluate the Lagrangian \eqref{lagdressed} on the ansatz \eqref{eq:ansatsvacuumcomplete}, obtaining
\begin{equation}\label{lagvuoto}
\begin{split}
\small
    \mathcal{L}_{\sigma}\left[\Sigma_0,\sigma_0\right]&=-e^{-4 f \sigma_{0} } \Lambda_0 ^4-V(\sigma_0)-\frac{R\ e^{-2 f \sigma_0 }}{12 f^2}+2 m^2_\pi \nu^2 X \cos\varphi\ e^{-f \sigma_{0}  y}\\&+ N_f \mu ^2 \nu^2 e^{-2 f \sigma_{0} } \sin ^2\varphi -a \nu^2 e^{-4 f \sigma_{0} } {\bar \theta}^2 \,.
\end{split}
\end{equation}
% where
% \be
% {\bar \theta} \equiv \theta -\sum_i^{N_f} \alpha_i \,, \qquad X \equiv \sum_i^{N_f} \cos \alpha_i  \ .
% \ee
It is worth noticing that, replacing $m_\pi \rightarrow \sqrt{2} \, m_\pi $ and $\mu \rightarrow \sqrt{2} \, \mu$, the resulting expression is identical to that found for 2-color QCD at finite baryon density, as explored in \cite{Bersini:2022bnx}. 
Nevertheless we further the results obtained in the previous work \cite{Bersini:2022bnx} by being able to solve for generic dilaton potentials \eqref{sigmapotDelta} and mass deformations rather than concentrating on specific values of $\Delta$ and $y$.

The classical ground state energy is computed by solving the following EOMs
% \begin{align}
%      \frac{\delta \mathcal{L}}{\delta \alpha}=\frac{\delta \mathcal{L}}{\delta \varphi}=\frac{\delta \mathcal{L}}{\delta \sigma_{0}}=0, \quad \frac{\delta \mathcal{L}}{\delta \mu}=\frac{Q}{V}\ ,
% \end{align}
% explicitly:
\begin{align}
\sin \varphi(N_f \mu ^2 e^{-2 f \sigma_0 } \cos \varphi -m_{\pi}^2 X e^{-f \sigma_0  y})&=0 \,,\\ \label{eqalphai}
a e^{-4 f \sigma_0 } \Bar{\theta}- m_{\pi}^2 \sin \alpha_i \cos \varphi  e^{-f \sigma_0  y}&=0 \,, \qquad i=1, .., N_f  \,, \\
 \frac{R e^{-2 f \sigma_0 }}{6f}+ 4 a f \nu ^2 e^{-4 f \sigma_0 } \Bar{\theta}^2+4 f \Lambda_{0}^4 e^{-4 f \sigma_0 }-\frac{\partial V(\sigma)}{\partial \sigma} {\bigg \rvert}_{\sigma=\sigma_0}+& \nonumber\\-2 f N_f \mu ^2 \nu^2 e^{-2 f \sigma_0 } \sin ^2\varphi 
    -2 f y m_{\pi}^2  \nu^2 X \cos \varphi  e^{-f \sigma_0  y}&=0  \,,\\\label{charge}
    2 N_f \mu \nu^2 e^{-2 f \sigma_0 } \sin ^2\varphi &=\frac{Q}{V}  \,,
\end{align}
where the last equation corresponds to fixing the isospin charge density. By solving the EOM via an expansion in the dilaton and pion masses, at the leading order in the double scaling limit \eqref{dsl}, we obtain 
{\small
\begin{equation}\label{deltagenerale}
\begin{split}
    \Delta_Q = r E_Q &= \frac{\pi^2}{8 f^2 } \left(  6 N_f(f \mu  \nu  r)^2 +1 \right) \left(\frac{2 N_f (f \mu  \nu  r)^2 -1}{f^2 \Lambda ^4 }\right)\  +m_{\sigma }^2\ \frac{ \pi ^2 2^{1-\Delta } r^{4-\Delta}}{(\Delta -4) \Delta  f^2 } \left(\frac{2 N_f (f \mu  \nu  r)^2 -1}{f^2 \Lambda ^4 }\right)^{\Delta /2}\   \\ & - m_\pi^4 N_f \cos^2  (\alpha(\theta)) \  2^{2 \gamma -3}   \left(\frac{\pi \nu  r^{\gamma+1}}{r \mu}\right)^2\left(\frac{2  N_f (f \mu  \nu  r)^2-1}{f^2 \Lambda ^4 }\right)^{2-\gamma }\    + \mathcal{O}\left ( m_\sigma^4 , m_\pi^8, m_\sigma^2 m_\pi^4\right)  \,,
\end{split}
\end{equation}}
where $\mu$ is related to $Q$ as
\begin{equation} \label{musol}
   r \mu  = \frac{\left(6 \pi ^4 \nu ^2 N_f\right)^{1/3}+\left(\sqrt{81 f^6 \Lambda ^8 Q^2-6 \pi ^4 \nu ^2 N_f}+9 f^3 \Lambda ^4 Q\right)^{2/3}}{f \left(6 \pi  \nu ^2 N_f\right)^{2/3} \left(\sqrt{81 f^6 \Lambda ^8 Q^2-6 \pi ^4 \nu ^2 N_f}+9 f^3 \Lambda ^4 Q\right)^{1/3}} \ .
\end{equation}
The first term in \eqref{deltagenerale} represents the scaling dimension in the conformal limit $m_\pi = m_\sigma = 0$ which depends only on the dimensionless combination $\mu r$ \eqref{musol} and, therefore, is  insensitive to the spacetime geometry. Noticeably, the leading correction in the pion mass is of order $m_\pi^4$ and its dependence on the geometry is tied to the anomalous dimension of the chiral condensate through the universal factor $r^{2(\gamma+1)}$. Remarkably, the whole $\theta$-dependence is encoded in the factor $\cos^2  (\alpha(\theta))$ with $\alpha(\theta)$ given in eq.\eqref{questaequazione}. The first term in the dilaton potential \eqref{sigmapotDelta} redefines the cosmological constant as
\be
\Lambda^4  \equiv \Lambda^4_0+\frac{m^2_\sigma}{4f^2(4-\Delta)}   \,.
\ee
The contribution stemming from the second term in the dilaton potential \eqref{sigmapotDelta} has been expanded in powers of $m_\sigma$ with the leading order  quadratic in the dilaton mass. The latter exhibits a universal dependence on the radius of the sphere via the factor $r^{4-\Delta}$. It is interesting to further expand our results  \eqref{deltagenerale} in the large charge limit $Q (\Lambda_0 f)^4 \gg 1$ where we can make contact with the universal predictions of the large charge effective field theory (EFT) \cite{Hellerman:2015nra, Gaume:2020bmp}. We have
\begin{equation}
\label{GSEnc2}
    \Delta_Q = r E_Q= \Delta_Q^\ast  +\left(\frac{m_{\sigma}}{4 \pi \nu}\right)^2 \, Q^{\frac{\Delta}{3}} B_1 +  \left(\frac{m_\pi}{4\pi \nu} \right)^4\ \cos^2  (\alpha(\theta))\  Q^{\frac{2}{3}(1-\gamma)} B_2  + \mathcal{O}\left ( m_\sigma^4 , m_\pi^8, m_\sigma^2 m_\pi^4\right)  \,,
     \end{equation}
where  $B_1$ and $B_2$ read
\begin{align}
B_1 & = \frac{c_{2/3} 2^{9-2 \Delta } 3^{\frac{\Delta }{2}-1}  ( \pi \nu  r)^{4-\Delta } \left(c_{4/3} N_f\right){}^{1-\frac{\Delta }{2}}}{(\Delta -4) \Delta } \left(1 -\frac{ \Delta\ c_{2/3} }{4 c_{4/3}}Q^{-2/3} +\cO\left(Q^{-4/3} \right) \right)  \,,\\
  B_2 &= - \ 3^{4-\gamma } 2^{4\gamma -3} \pi ^{2 \gamma +2} c_{4/3}^{\gamma -4} N_f^{\gamma -1} (\nu  r)^{2 (\gamma +1)}\left(1 +\frac{ (\gamma -4) c_{2/3}}{2 c_{4/3}} Q^{-2/3} +\cO\left(Q^{-4/3}\right)\right) \,,
\end{align}
%\tad{$B_2$ non torna}
while
    \be \label{confdim}
\Delta_Q^\ast =r E_Q= c_{4/3} Q^{4/3} +c_{2/3} Q^{2/3} + \cO\left(Q^0 \right) \,,
   \ee
    is the scaling dimension in the conformal limit which depends only on the dimensionless coefficients defined below 
\begin{equation} \label{coeffic}
    c_{4/3} =\frac{3}{8}\left(\frac{2 \Lambda^2}{\pi N_f \nu^2}\right)^{2/3}\,, \quad c_{2/3} =\frac{1}{4 f^2}\left(\frac{2 \pi^2}{N_f \nu^2 \Lambda^4}\right)^{1/3}\,.
\end{equation}
The conformal dimension $\Delta_Q^\ast$ exhibits the general structure predicted by the large charge EFT. The non-conformal corrections feature a characteristic $Q$-scaling which has been made manifest in eq.\eqref{GSEnc2} and depends on the parameters $\gamma$ and $\Delta$ encoding the explicit breaking of scale invariance. 

\section{SSB and spectrum of fluctuations}
\label{sectionfluc}
We now move to determine the spectrum of the theory including the symmetry breaking pattern starting from the latter. In fact, fixing the generalized isospin charge results in the following symmetry breaking pattern
{\small
\begin{align}
   SU(N_f)_L \times SU(N_f)_R  \times U(1)_V\overset{N_f^2-1}{\rightsquigarrow} \ & SU(N_f)_V \times U(1)_V \longrightarrow SU\left(\frac{N_f}{2}\right)_u \times SU\left(\frac{N_f}{2}\right)_d \times U(1)_I \times U(1)_V \nonumber\\&\overset{\frac{N_f^2}{4}}{\rightsquigarrow} SU\left(\frac{N_f}{2}\right)_{ud} \times U(1)_V   \,,
\end{align}}
where $\rightsquigarrow$ and $\longrightarrow$ denote, respectively, spontaneous and explicit breaking.   The first stage is the usual chiral symmetry breaking with the ABJ anomaly already taken into account in the breaking of the axial symmetry. The further explicit breaking is owed to the introduction of the isospin charge while the last spontaneous breaking is associated with pion condensation and the superfluid phase transition. In the absence of the dilaton, the spectrum of light modes is composed of $N_f^2/4$ massless Goldstone bosons with speed $v_G=1$ that parameterize the coset $G/H=\displaystyle{\frac{SU(N_f/2)_u \times SU(N_f/2)_d \times U(1)_I \times U(1)_V}{SU(N_f/2)_{ud} \times U(1)_V }}$. These modes arrange themselves in the adjoint representation of the stability group $SU(N_f/2)_{udV} \times U(1)_V$ plus a singlet which we denote as $\pi_3$ since it is associated with the spontaneous breaking of $U(1)_I$ i.e. to the third Pauli matrix in the $N_f=2$ case. In addition, a pseudo-Goldstone mode stems from the would-be spontaneous breaking of $U(1)_A$ which we call the $S$ (singlet) mode and it is related to the $\eta^\prime$-meson \cite{DiVecchia:2013swa}.  As mentioned above, the $U(1)_A$ symmetry is quantum mechanically anomalous, and therefore the latter mode acquires a mass proportional to the scale of the anomaly factor $\sqrt{a}$.

In what follows, we shall focus on the spectrum of Goldstone bosons since they control the large charge dynamics. Specifically, we are interested in analyzing how the spectrum of light modes changes when (near)conformal dynamics is realized through the dilaton dressing. Precisely, conformal invariance dictates the existence of a massless mode with speed $v_G = \frac{1}{\sqrt{d-1}}= \frac{1}{\sqrt{3}}$ \cite{Hellerman:2015nra, Son:2002zn}. 
As we shall see, the latter arises from the mixing between the singlet $\pi_3$ with the dilaton that acts as its "radial mode" and changes its speed from $v_G =1$ to $v_G= \frac{1}{\sqrt{3}}$.  We consider the hierarchy of scales $m_\pi, m_\sigma \ll  \mu \ll 4 \pi \nu$ which ensures the validity of chiral perturbation theory and assumes small deviation from conformality. To determine the fluctuations' spectrum, we expand around the vacuum solution as
\begin{align}
\label{perttheta}
   \Sigma = e^{\frac{2i}{\sqrt{N_f}} S \mathbbm{1}_{N_f}}e^{i\Omega}\Sigma_0 e^{-i\Omega^\dagger}  \,, 
\end{align}
where $\Sigma_{0}$ is the classical solution \eqref{eq:ansatsvacuum} while the fluctuations are organized in the matrix $\Omega$ as 
\begin{align}
 \Omega= \left(\begin{array}{cc}
\pi  & 0 \\
0 & -\pi^{t}  \\
\end{array}\right) \,.
\end{align}
%devo verificare che la normalizzazione possa essere diversa per il generatore di U(1)A e per gli altri generatori rotti pper cui dovrebbe invece essere 1/sqrt(2s) !!!
Here $\pi=\sum_{a=0}^{\text{dim G/H}} \pi^{a}T_{a}$ belongs to the algebra of the coset space  and we normalized the generators as $Tr\left\{ T_{a}T_{b}\right\} =\frac{\delta_{ab}}{2}$. After some manipulations, we obtain  
\begin{align}
    Tr\left\{\partial_{\mu} \Sigma \partial^{\mu} \Sigma^{\dagger}\right\} &=4\sin^2\varphi\ \partial_{\mu}\pi^{a}\partial^{\mu}\pi^{a}+4 \partial_{\mu}S\partial^{\mu}S   \,, \\
    Tr\left\{ I\partial_0 \Sigma \Sigma^\dagger - I\Sigma^\dagger \partial_0 \Sigma \right\} &=4i Tr\left\{ I \partial_0 \nu \right\} \sin^2 \varphi = 2i \sqrt{N_f} \, \partial_0 \pi^3 \sin^2 \varphi  \,, \\
    Tr\left\{ M\Sigma+M^{\dagger}\Sigma^{\dagger}\right\} &=2\cos\varphi\left[X \cos \left( \frac{2}{\sqrt{N_f}} S\right)+Z \sin \left(\frac{2}{\sqrt{N_f}} S\right)\right]  \,,\\
    Tr\left\{ \log \Sigma-\log \Sigma^{\dagger}\right\}&=4i \sqrt{N_f} S-2i\sum_i^{N_f}\alpha_i\ ,
\end{align}
where we defined $Z \equiv \sum_{i=1}^{N_f} \sin \alpha_i$. 
Finally, by expanding the dilaton field around its background solution as $\sigma\rightarrow\sigma_{0}+\hat{\sigma}(t,\mathbf{x})$, we arrive at the following quadratic Lagrangian
\begin{align}
\label{ldivkin}
\frac{\mathcal{L}_\sigma^{(2)}}{4\nu^2\sin^2\varphi\ e^{-2\sigma_{0}f}}=\left(\begin{array}{lll}
\pi^3 & \hat{\sigma} &S
\end{array}\right)D^{-1}\left(\begin{array}{l}
\pi^3 \\
\hat{\sigma} \\
S
\end{array}\right) + \sum_{a=1, a\neq3}^{\text{dim}[{\tiny\yng(2,1)}]} \partial_\mu \pi^a \partial^\mu \pi^a  \,,
\end{align}
where the inverse propagator $D^{-1}$ is given by
\begin{align}
\label{d-1theta}
  D^{-1}&=\left(\begin{array}{ccc}
\omega^{2}-k^{2} & i\omega \mu f \sqrt{N_f} & 0 \\
-i\omega \mu f \sqrt{N_f} & \frac{\omega^{2}-k^{2}}{8\nu^2\sin^2\varphi}-M_{\sigma}^{2} & \frac{ 1}{2} I_{\hat{\sigma}s} \\
0 &\frac{1}{2} I_{\hat{\sigma}s} &\frac{\omega^{2}-k^{2}}{\sin^2\varphi}-M_{s}^{2}
\end{array}\right)\ , 
\end{align}
with
\begin{align}
    I_{\hat{\sigma} S}&=\frac{ f \mu ^2 \sqrt{N_f} \left(4 a \mu ^2  \Bar{\theta} N_f^2 e^{-2 f \sigma _0 \gamma}-m_{\pi }^4 X  Z(3-\gamma)\right)}{ \left( \mu ^4 N_f^2 e^{2 f \sigma _0 (1-\gamma)}-m_{\pi }^4 X^2\right)}  \,, \\
    M_\sigma^2 & = \frac{\mu ^2 N_f }{8 \nu ^2 }\Bigg[\frac{\Delta  \mu ^2 N_f m_{\sigma }^2 e^{-f(\Delta-2 )  \sigma }}{(\Delta -4) \left(\mu ^4 N_f^2-m_{\pi }^4 X^2 e^{2 (\gamma -1) f \sigma }\right)}\nonumber\\&+\left.\frac{2 f^2 \left(2 \mu ^2 N_f \left(\mu ^2 \nu ^2 N_f e^{2 f \sigma }-4 \left(a \nu ^2 \Bar{\theta}^2+\Lambda ^4\right)\right)+((\gamma -6) \gamma +7) \nu ^2 m_{\pi }^4 X^2 e^{2 \gamma  f \sigma }\right)}{m_{\pi }^4 X^2 e^{2 \gamma  f \sigma }-\mu ^4 N_f^2 e^{2 f \sigma }}\right)\Bigg]  \,, \\ \label{massofS}
   M_S^2 &= \frac{a \mu ^4 N_f^3 +\mu ^2 m_\pi^4 X^2 e^{2 \gamma  f \sigma_0 }}{\mu ^4 N_f^2 e^{2 f \sigma_0 }-m_\pi^4 X^2 e^{2 \gamma  f \sigma_0 }}
\ .
\end{align}
The $\pi_a$ modes with $a \neq 3$ appearing in eq.\eqref{ldivkin} denote the Goldstone modes transforming under the adjoint representation of $SU(N_f/2)$ and have trivial dispersion relations $\omega = k$. On the other hand, $\pi_3$ transforms as a singlet of $SU(N_f/2)$, and due to this property, it mixes with the dilaton and the $S$, as it is manifest in eq.\eqref{d-1theta}. We pause to note that eq.\eqref{massofS} is consistent with the Witten-Veneziano relation \cite{Witten:1979vv, Veneziano:1979ec}. In fact, in the $m_\pi \to 0$ limit,$ M_S^2$ reduces to
\begin{equation} \label{WV}
    \lim_{\sigma_0\to 0,\ m_\pi\to 0} \eqref{massofS} \implies M^2_{S}=a N_f\ .
\end{equation}

The remaining dispersion relations are obtained by solving the equation det$(D^{- 1})$. The results describe two gapped modes $\omega_{1,2}$ with mass
{\small
\begin{align}   
    M^2_{1,2}&=-\frac{1}{2} \sin^2\varphi \left[M_S^2+8 \nu ^2 f^2 \mu ^2 N_f+8 \nu ^2 M_{\sigma }^2 \pm \sqrt{\left(M_S^2-8 \nu ^2 \left(f^2 \mu ^2 N_f+M_{\sigma }^2\right)\right){}^2 + 8 \nu ^2 I_{\hat{\sigma} S}^2}\right]  \,,
\end{align} 
}
and a massless mode $\omega_3$ with speed given by
\begin{align}
    v_3^2&=\frac{I_{\hat{\sigma}s}^2-4 M^2_\sigma M^2_S}{I_{\hat{\sigma}s}^2-4 M^2_S\left(M^2_\sigma+ f^2 N_f \mu^2\right)}\ .
\end{align}
In parallel with the previous section, we determine the dispersion relations considering corrections in both $m_\sigma^2$ and $m_\pi^2$. According to this expansion, we obtain
\begin{align}\label{masslessomega}
\omega^2_{i}&=\omega^{2\ast}_{i}+ \left(\frac{N_f \mu^2 \nu^2}{2\Lambda ^4}\right)^{\frac{\Delta -2}{2}}m^2_\sigma D_1(\omega_i^{2\ast}, \Delta) +\left(\frac{N_f \mu^2 \nu^2}{2\Lambda ^4}\right)^{-\gamma-3} m^4_\pi  \cos^2(\alpha(\theta)) D_2(\omega_i^{2\ast},  \gamma) \nonumber \\ &  + \mathcal{O}\left ( m_\sigma^4 , m_\pi^8, m_\sigma^2 m_\pi^4\right)  \,, \qquad  \text{for} \ i=1,2,3 \,.
\end{align}
The coefficients $D_1$ and $D_2$ read 
{\footnotesize
    \begin{align}
        D_1(\omega_i^{2\ast}, \Delta) &= \frac{1}{12 (\Delta -4) f^2 \left(\mu ^2 \nu ^2 N_f \left(a N_f \left(6 f^2 \mu ^2 \nu ^2 N_f+k^2-\omega_i^{2\ast} \right)+8 f^2 \Lambda ^4 \left(2 k^2-3 \omega_i^{2\ast} \right)\right)+3 \Lambda ^4 \left(k^2-\omega_i^{2\ast} \right)^2\right)}\nonumber\\& \times\left[a N_f \left(2 f^2 \mu ^2 \nu ^2 N_f \left((3 \Delta -10) k^2-3 (\Delta -2) \omega_i^{2\ast} \right)-\left(k^2-\omega_i^{2\ast} \right)^2\right)\right.\nonumber\\&\left. +4 f^2 \Lambda ^4 \left(k^2-\omega_i^{2\ast} \right) \left((3 \Delta -8) k^2-3 \Delta  \omega_i^{2\ast} \right)\right] \,,\\
        D_2(\omega_i^{2\ast}, \gamma) &= \frac{\mu ^4 \nu ^8 \Lambda ^{-4 (\gamma +4)} (\mu  \nu )^{2 \gamma } N_f^{\gamma +4}}{96 \left(\mu ^2 \nu ^2 N_f \left(a N_f \left(k^2-\omega_i^{2\ast} \right)+8 f^2 \Lambda ^4 \left(2 k^2-3 \omega_i^{2\ast} \right)\right)+6 a f^2 \mu ^4 \nu ^4 N_f^3+3 \Lambda ^4 \left(k^2-\omega_i^{2\ast} \right)^2\right)}\nonumber\\ & \times \left[ -2 (\gamma  (3 \gamma -10)-5) f^2 \Lambda ^{4 \gamma } \mu ^{2-2 \gamma } \left(k^2-\omega_i^{2\ast} \right) \left(\nu ^2 N_f\right)^{1-\gamma } \left(a \mu ^2 \nu ^2 N_f^2+2 \Lambda ^4 \left(k^2-\omega_i^{2\ast} \right)\right)\right. \nonumber\\&\left. +3 \left(k^2-\omega_i^{2\ast} \right)^2 \left(\frac{\mu ^2 \nu ^2 N_f}{\Lambda ^4}\right)^{-\gamma } \left(a \mu ^2 \nu ^2 N_f^2+2 \Lambda ^4 \left(k^2-\omega_i^{2\ast} \right)\right)-16 (\gamma -1) f^2 \omega_i^{2\ast}  \Lambda ^{4 \gamma } \mu ^{2-2 \gamma } \left(\nu ^2 N_f\right)^{1-\gamma }\right.\nonumber\\&\left. \times \left(a \mu ^2 \nu ^2 N_f^2+2 \Lambda ^4 \left(k^2-\omega_i^{2\ast} \right)\right)+\Lambda ^{4 \gamma } (\mu  \nu )^{-2 \gamma } N_f^{-\gamma } \left(4 f^2 \mu ^2 \nu ^2 N_f \left(k^2-3 \omega_i^{2\ast} \right)+\left(k^2-\omega_i^{2\ast} \right)^2\right)\right.\nonumber \\& \times \left. \left(6 \Lambda ^4 \left(k^2+\mu ^2-\omega_i^{2\ast} \right)-a (\gamma -7) \mu ^2 \nu ^2 N_f^2\right)\right]
    \end{align}
    }
while the dispersion relations in the conformal limit $m_\pi=m_\sigma=0$ have the following simple form
\begin{align}
    \omega^{2\ast}_{1} & = k^2 +6 f^2 \mu ^2 \nu ^2 N_f+ 2 f \mu  \nu  \sqrt{N_f \left(9 f^2 \mu ^2 \nu ^2 N_f+2 k^2\right)}  \,,\\
    \omega^{2\ast}_{2} & = k^2+ \frac{a \mu ^2 \nu ^2 N_f^2}{2 \Lambda^4}  \,,\\
      \omega^{2\ast}_{3} &= k^2 +6 f^2 \mu ^2 \nu ^2 N_f -2 f \mu  \nu  \sqrt{N_f \left(9 f^2 \mu ^2 \nu ^2 N_f+2 k^2\right)}  \,,
\end{align}

        % D_2(\omega_i^{2\ast}, \gamma) &= \frac{ \mu ^4 \nu ^8 N_f^2  }{96 \Lambda ^{16} \left(\mu ^2 \nu ^2 N_f \left(a N_f \left(k^2-\omega_i^{2\ast} \right)+8 f^2 \Lambda ^4 \left(2 k^2-3 \omega_i^{2\ast} \right)\right)+6 a f^2 \mu ^4 \nu ^4 N_f^3+3 \Lambda ^4 \left(k^2-\omega_i^{2\ast} \right)^2\right)}\nonumber\\& \times  \Bigg[3 \left(k^2-\omega_i^{2\ast} \right)^2 \left(2 \Lambda ^4 \left(k^2-\omega_i^{2\ast} \right)+a \mu ^2 \nu ^2 N_f^2\right)-2 (3 \gamma^2 -10\gamma-5) f^2  \mu^2 \left(k^2-\omega_i^{2\ast} \right) \nu ^2 N_f \nonumber\\& \times \left(a \mu ^2 \nu ^2 N_f^2+2 \Lambda^4 \left(k^2-\omega_i^{2\ast} \right)\right)  -16 (\gamma -1) f^2 \omega_i^{2\ast}   \mu^2 \nu ^2 N_f \left(a \mu ^2 \nu ^2 N_f^2+2 \Lambda ^4 \left(k^2-\omega_i^{2\ast} \right)\right) \nonumber\\&  +2 \left(4 f^2 \mu ^2 \nu ^2 N_f \left(k^2-3 \omega_i^{2\ast} \right)+\left(k^2-\omega_i^{2\ast} \right)^2\right)  \left(3 \Lambda^4 \left(k^2+\mu ^2-\omega_i^{2 \ast} \right)-\frac{a}{2} (\gamma -7) \mu ^2 \nu ^2 N_f^2\right)\Bigg]\ ,
where we can recognize the expected mode with a square mass of order $a$ stemming from the axial anomaly as well as a gapped mode with mass $12 N_f f^2 \mu^2 \nu^2$ and a Goldstone boson with speed $v_3^2 =\frac{1}{3} $ as dictated by scale invariance. We conclude the section by providing explicit expression for $M_{1,2}^2$ and $v_3^2$

\allowdisplaybreaks
{\small
\begin{align}
    v^2_{3} &= \frac{1}{3}+ m^2_\sigma\  \left(\frac{N_f \mu^2 \nu^2}{2\Lambda ^4}\right)^{\Delta/2}\frac {\Lambda ^4}{9 f^2 N_f^2 \mu^4 \nu^4}\nonumber\\& -m^4_\pi \cos^2  (\alpha(\theta))  \left(\frac{N_f \mu^2 \nu^2}{2\Lambda ^4}\right)^{-\gamma-3}\frac{ \mu ^4 \nu ^8  N_f^{2}}{144 \Lambda ^{16}} (\gamma -3) (\gamma +1)  + \mathcal{O}\left ( m_\sigma^4 , m_\pi^8, m_\sigma^2 m_\pi^4\right)  \,,\\
    M^2_{1} &= 12 N_f f^2 \mu^2 \nu^2 + m^2_\sigma\  \left(\frac{N_f \mu^2 \nu^2}{2\Lambda ^4}\right)^{\Delta/2} \frac{ \Delta  }{\Delta -4}\left(\frac{2\Lambda ^4}{N_f \mu^2 \nu^2}\right)\nonumber\\ &-m^4_\pi \cos^2  (\alpha(\theta))\   \left(\frac{N_f \mu^2 \nu^2}{2\Lambda ^4}\right)^{-\gamma-3} \frac{f^2 \mu ^6 \nu ^{10}  N_f^3}{8 \Lambda ^{16}} ( \gamma^2 -6\gamma +7)  + \mathcal{O}\left ( m_\sigma^4 , m_\pi^8, m_\sigma^2 m_\pi^4\right)  \,, \\
    M^2_{2} &= \frac{a \mu ^2 \nu ^2 N_f^2}{2 \Lambda^4} - m^2_\sigma\ \left(\frac{N_f \mu^2 \nu^2}{2\Lambda ^4}\right)^{\Delta/2} \frac{a}{6 (\Delta -4) f^2 \mu^2 \nu^2} \nonumber\\& -m^4_\pi \cos^2  (\alpha(\theta))\  \left(\frac{N_f \mu^2 \nu^2}{2\Lambda ^4}\right)^{-\gamma-3}\frac{ \mu ^6 \nu ^8   N_f^2 }{96 \Lambda^{20}}  \left(a (\gamma -4) \nu ^2 N_f^2-6 \Lambda ^4\right)\ + \mathcal{O}\left ( m_\sigma^4 , m_\pi^8, m_\sigma^2 m_\pi^4\right)  \,.
\end{align}
}
%\tad{Controlla fattori $N_f^2$ da $X$ in correzioni proporzionali alla pion mass} 

\section{Casimir energy contribution to $\Delta_Q$ in the conformal limit}
\label{Casimiren}
As discussed in Sec.\ref{DACL}, in the conformal limit $m_\pi= m_\sigma= 0$ our results \eqref{GSEnc2} correspond to the leading contribution to the conformal dimension $\Delta_Q$ of the lowest-lying charge $Q$ operator in the double scaling limit \eqref{dsl}. However, as has been first shown in \cite{Hellerman:2015nra}, $\Delta_Q$ can also be computed in the strongly coupled regime by constructing an EFT for the relativistic Goldstone modes stemming from the SSB induced by fixing the charge. In $d=4$ dimensions, the large charge EFT predicts \cite{Hellerman:2015nra, Cuomo:2020rgt} 
\be \label{confindustria}
\Delta_Q = k_{4/3} Q^{4/3} +k_{2/3} Q^{2/3} + k_0 \log Q + \cO\left(Q^0 \right) \,,
\ee
where the coefficients $k_{4/3}$ and $k_{2/3}$ are related to the Wilson coefficients of the EFT and cannot, therefore, be computed within the EFT approach. The calculated coefficients $c_{4/3}$ and $c_{2/3}$ given in eq.\eqref{coeffic} can be seen as the leading contribution to $k_{4/3}$ and $k_{2/3}$ in a perturbative expansion of the latter around $(\Lambda_0 f)^4=0$ \cite{Badel:2019oxl}. On the other hand, the coefficient $k_0$ is a purely quantum contribution related to the Casimir energy of the relativistic Goldstone bosons. Importantly, its value is universal being entirely determined by symmetry and the number of spacetime dimensions. In particular, it can be computed exactly from the knowledge of the low energy spectrum. In fact, consider the low-energy action for a Goldstone mode $\chi$
\begin{equation}
    S_G=\int_{\mathcal{M}}\ \dd t \ \dd \mathbf{x}\ \left(\frac{1}{2}(\partial_t\chi)^2+\frac{v^2_G}{2}\big(\grad{\chi}\big)^2\right)\ .
\end{equation}
The corresponding Casimir energy is given by
\begin{equation}
\label{flucdet}
    E_{\rm Casimir}=\frac{1}{2}Tr\{\log\big(-\partial^2_t-v^2_G \laplacian\big)\} = \frac{1}{4\pi}\int_{-\infty}^{\infty}\ \dd\omega \sum_{\mathbf{p}}\ \log\big(\omega^2+v^2_G E^2(\mathbf{p})\big) = \frac{v_G}{2} \sum_{\mathbf{p}} E(\mathbf{p})  \,.
\end{equation}
Here $E(\mathbf{p})^2$ denotes the eigenvalues of the Laplacian operator on $S^{3}$. The above contribution scales as $Q^0$ and exhibits a pole for $d \to 4$ in dimensional regularization. The latter is related to a   $\log Q$ term with a universal coefficient $-\frac{v_G}{48}$ stemming from the renormalization of the vacuum energy \cite{Cuomo:2020rgt}. Hence, to calculate $k_0$ we simply need to sum the contributions of the various Goldstone bosons. The symmetry breaking pattern in the chiral limit reads
% \begin{align}
%     SU(N_f)_L \times & SU(N_f)_R \times U(1)_V \times U(1)_A \overset{N_f^2}{\leadsto} SU(N_f)_V \times U(1)_V \times U(1)_A \rightarrow  \nonumber\\ & \rightarrow SU\left(\frac{N_f}{2}\right)_{uL} \times SU\left(\frac{N_f}{2}\right)_{dL} \times SU\left(\frac{N_f}{2}\right)_{uR} \times SU\left(\frac{N_f}{2}\right)_{dR} \times  U(1)_I \times U(1)_V  \nonumber \\&  \overset{\frac{N_f^2}{2}-1}{\leadsto} SU\left(\frac{N_f}{2}\right)_{udL} \times SU\left(\frac{N_f}{2}\right)_{udR} \times U(1)_V  \,.
% \end{align}
\begin{align}
   & SU(N_f)_L \times  SU(N_f)_R \times U(1)_V \times U(1)_A   \nonumber\\ & \rightarrow SU\left(\frac{N_f}{2}\right)_{uL} \times SU\left(\frac{N_f}{2}\right)_{dL} \times SU\left(\frac{N_f}{2}\right)_{uR} \times SU\left(\frac{N_f}{2}\right)_{dR} \times  U(1)_I \times U(1)_V  \nonumber \\&  \overset{\frac{3}{4}N_f^2-2}{\leadsto} SU\left(\frac{N_f}{2}\right)_{ud} \times U(1)_V  \,.
\end{align}
The resulting $\frac{3}{4}N_f^2-2$ Goldstone bosons have speed $v_G = 1$ except the $\pi^3$ mode which has $v_G = \frac{1}{\sqrt{3}}$. We therefore obtain 
\begin{equation}
   k_0=-\frac{1}{48}\left(\frac{1}{\sqrt{3}}+ \frac{3}{4} N_f^2-3\right) .
\end{equation}

\section{Conclusions}
\label{Conclusions}

We uncovered near-conformal properties of finite isospin density  QCD on a non-trivial gravitational background. Specifically, we determined the ground state energy  $\Delta_Q/r$ with $r$ radius of $\mathbb{R}\times S^3$ via the semiclassical large charge expansion. In the conformal limit this energy maps, via state-operator correspondence, into the scaling dimension $\Delta_Q^\ast$ of the lowest lying operator of fixed isospin charge $Q$. 

One of our main results given in \eqref{Alessandra} is the determination of the leading near conformal corrections to $\Delta_Q^\ast$  at the lower boundary of the QCD conformal window.  We showed that the characteristic $Q$-scalings due to the near conformal corrections are induced by the quark mass operator anomalous dimension $\gamma$ as well as the conformal dimension $\Delta$ of the operator responsible for dynamically deforming QCD away from the conformal window. Our results and methodology work as a template to obtain similar results for QCD-like theories such as two-color QCD at nonzero baryon density.

Additionally, we determined the pattern of symmetry breaking and the associated physical spectrum and their dispersion relations.  These latter results will also help determine the next-to-leading order large charge contributions.  Last but not least we discussed the $\mu-\theta$ QCD phase diagram.

\section*{Acknowledgments}

The work of J.B. was supported by the World Premier International Research Center Initiative (WPI Initiative), MEXT, Japan; and also supported by the JSPS KAKENHI Grant Number JP23K19047. The work of F.S.~is partially supported by the Carlsberg Foundation, grant CF22-0922. A.D.A. expresses sincere appreciation to the University of Southern Denmark and D-IAS for their hospitality during the crucial final stages of the work.

% \newpage
%\begin{appendices}
%    \section{On the details of computations}  
%\label{appendixa}

%    \end{appendices}

%----------NEW BIB--------------
%\addcontentsline{toc}{chapter}{Bibliography}
%\bibliographystyle{plain}
%\bibliography{bibliography}
%------------------------------- 

\end{document}